\documentclass{iopart}
\usepackage{iopams}
\usepackage{bbm}
\usepackage{graphicx}
\usepackage{color}
\usepackage{stmaryrd}

\newcommand{\me}[3]{\ensuremath{\langle#1|#2|#3\rangle}}
\newcommand{\avg}[1]{\ensuremath{\langle#1\rangle}}
\newcommand{\scp}[2]{\ensuremath{\langle#1|#2\rangle}}

\newcommand{\binom}[2]{\ensuremath{\left(\begin{array}{c}#1\\#2\end{array}\right)}}

\newcommand{\dagg}{\ensuremath{^{\dagger}}}

\newcommand{\op}[1]{\ensuremath{\hat{#1}}}

\newcommand{\se}[1]{\ensuremath{^{\mathrm{#1}}}}
\newcommand{\si}[1]{\ensuremath{_{\mathrm{#1}}}}

\definecolor{bulletcolor}{RGB}{0,139,0}

\begin{document}

\title{Tomographic reconstruction of the Wigner function on the Bloch sphere}
\author{Roman Schmied and Philipp Treutlein}
\address{Departement Physik, Universit\"at Basel, Klingelbergstrasse 82, CH--4056 Basel, Switzerland}
\ead{roman.schmied@unibas.ch}

\begin{abstract}
We present a filtered backprojection algorithm for reconstructing the Wigner function of a system of large angular momentum $j$ from Stern--Gerlach-type measurements. Our method is advantageous over the full determination of the density matrix in that it is insensitive to experimental fluctuations in $j$, and allows for a natural elimination of high-frequency noise in the Wigner function by taking into account the experimental uncertainties in the determination of $j$, its projection $m$, and the quantization axis orientation. No data binning and no arbitrary smoothing parameters are necessary in this reconstruction. Using recently published data [Riedel \emph{et al.}, Nature \textbf{464}:1170 (2010)] we reconstruct the Wigner function of a spin-squeezed state of a Bose--Einstein condensate of about 1250 atoms, demonstrating that measurements along quantization axes lying in a single plane are sufficient for performing this tomographic reconstruction. Our method does not guarantee positivity of the reconstructed density matrix in the presence of experimental noise, which is a general limitation of backprojection algorithms.
\end{abstract}

\section{Introduction}
\label{sec:intro}

The reconstruction of the quantum-mechanical state of a system from measurements is an important topic of the emerging field of quantum technology~\cite{QuantumStateEstimation}. Through partial or full state reconstruction we can estimate entanglement properties of multipartite quantum systems, and judge their usefulness for further experimental progress in fields such as quantum metrology~\cite{Treutlein2004,Esteve2008,Appel2009,Gross2010,Riedel2010,Leroux2010b,SchleierSmith2010,Leroux2010a}, quantum simulation~\cite{Friedenauer2008}, and quantum computation~\cite{Sackett2000,Leibfried2005,Haeffner2005,Treutlein2006b,Benhelm2008}.

Particularly in quantum metrology, experiments often involve large numbers of particles, and single-particle resolution is unavailable in both control and measurement. Because of this limitation, standard methods for reconstructing the quantum-mechanical density matrix~\cite{Newton1968,Klose2001,Haeffner2005} cannot be applied. For instance, and centrally to this work, in a Bose--Einstein condensate consisting of $N$ atoms, with each atom representing a pseudo-spin-$1/2$ subsystem, the total spin length $j=N/2$ can take on very large values and the known reconstruction procedures become problematic. In a single Stern--Gerlach measurement on the atomic ensemble we measure the numbers of up and down spins $N_{\uparrow}$ and $N_{\downarrow}$, in terms of which the total spin is $j=(N_{\uparrow}+N_{\downarrow})/2$ and the projection quantum number is $m=(N_{\uparrow}-N_{\downarrow})/2$. Since it is very difficult to determine the populations $N_{\uparrow}$ and $N_{\downarrow}$ with atomic accuracy~\cite{Buecker2009,Ockeloen2010}, the density matrix, which requires knowledge of $j$, becomes impossible to reconstruct in full. Further, reconstructing the $(2j+1)^2$ degrees of freedom of the density matrix~\cite{Newton1968,Klose2001,James2001} requires at least as many uncorrelated measurements, and therefore the experimental uncertainty in $m$ will hinder this full determination. In the absence of reliable data, there will be significant uncertainty and noise throughout the density matrix in its Dicke representation $\rho_{m m'}=\me{j m}{\op{\rho}}{j m'}$, which severely limits its usefulness. We need a method for calculating those components of $\op{\rho}$ which are significant even in the presence of noise and for very large values of $j$, and a way of determining which components must remain unknown.

The Wigner function~\cite{Schleich} is ideal for such a controlled reconstruction. It is a real-valued function on a sphere of radius $\hbar\sqrt{j(j+1)}$, represented in terms of orthonormal Laplace spherical harmonics as~\cite{Dowling1994}
\begin{equation}
	\label{eq:Wsph}
	W(\vartheta,\varphi) = \sum_{k=0}^{2j} \sum_{q=-k}^k \rho_{k q} Y_{k q}(\vartheta,\varphi),
\end{equation}
where $\vartheta$ is the polar angle measured from the $+z$ axis, and $\varphi$ is the azimuthal angle around the $z$ axis. While this sphere is commonly called a generalized Bloch sphere~\cite{Appel2009}, its surface actually represents a two-dimensional phase space instead of a Hilbert space as for the original Bloch sphere. This Wigner function contains the same information as the density matrix for any spin-$j$ system. While the marginals of the better-known Wigner function in planar space~\cite{Schleich,Vogel1989,Smithey1993,Breitenbach1997} are real-space or momentum-space probability distributions, the marginals of the spherical Wigner function are the projection quantum number distributions along all quantization axes [see~\eref{eq:projprob} below]; further, the expectation value of the angular momentum vector is proportional to the ``center of mass'' of the Wigner function, $\{\langle S_x \rangle,\langle S_y \rangle,\langle S_z \rangle\} = \sqrt{\frac{j(j+1)(2j+1)}{4\pi}}\times\int_0^{\pi}\sin\vartheta\rmd\vartheta \int_0^{2\pi}\rmd\varphi \{\sin\vartheta\cos\varphi, \sin\vartheta\sin\varphi,\cos\vartheta\} W(\vartheta,\varphi)$.

Most importantly, the Wigner function allows us to differentiate between more significant components $\rho_{k q}$ (with smaller values of $k$) and more noise-prone components (with larger values of $k$) in a natural way. Further, if only components with $k\ll2j$ are reconstructed, then accurate knowledge of $j$ is not necessary. As detailed in \sref{sec:math}, the transformation from $j$-space (the Dicke representation $\rho_{m m'}$ of the density matrix) to $k$-space (the spherical harmonic decomposition $\rho_{k q}$ of the Wigner function) proceeds though coupling coefficients which, at low $k$, are smooth in both $j$ and $m$; this significantly reduces the impact of uncertainties in the experimental determination of $(j,m)$.

Methods for reconstructing \emph{planar} Wigner functions by inverse Radon transform are well established in the context of nonlinear optics~\cite{Smithey1993,Breitenbach1997}. In the past they have also been applied to tomographic data on large-spin quantum systems, locally approximating the Bloch sphere by a tangental plane and neglecting its curvature~\cite{Riedel2010}. While this approximation is valid for spin states which are very localized on the Bloch sphere and do not wrap around it, future experimental progress is expected to produce ever more delocalized states (\emph{e.g.}, Schr\"odinger-cat states) whose properties are strongly influenced by the spherical shape of the Bloch sphere. Previous work on the reconstruction of the Wigner function on the full Bloch sphere has used the Husimi-$Q$ distribution as input~\cite{Agarwal1998}, which is the convolution of the system's Wigner function with that of a coherent state (see \sref{sec:planar}). This convolution washes out features of the Wigner function that are smaller than a coherent state. Since the principal characteristic of spin-squeezed states is that their Wigner function possesses a peak width smaller than that of a coherent state, such a deconvolution-based reconstruction approach is ill suited for studying spin-squeezed states, which is the goal of much current research in atomic physics~\cite{Treutlein2004,Esteve2008,Appel2009,Gross2010,Riedel2010,Leroux2010b,SchleierSmith2010,Leroux2010a}. We therefore require a new method for reconstructing the complex quantum-mechanical states of large-spin systems from experimental data in the absence of simplifying circumstances, such as strong phase-space localization and/or lack of spin squeezing.

This paper is organized as follows. In \sref{sec:math} we present a novel filtered backprojection algorithm for reconstructing the Wigner function from experimental Stern--Gerlach data. \Sref{sec:planar} specializes this algorithm to data acquired with quantization axes lying in a single plane. Finally, \sref{sec:data} applies the latter algorithm to a data set acquired in our group~\cite{Riedel2010}.
In what follows, a ``single Stern--Gerlach measurement'' describes a single determination of the projection quantum number $m$ of a quantum system along a certain quantization axis. In our case this corresponds to a single run of state preparation and population determination of a two-component Bose--Einstein condensate, yielding a single tuple $(j_n,m_n)$. The equivalent for the original experiment~\cite{Gerlach1922} is sending a single silver atom through the experimental apparatus, and determining its deflection by the magnetic field gradient. On the other hand, a ``Stern--Gerlach experiment'' is a series of many single Stern--Gerlach measurements with fixed quantization axis, sufficient to determine the probability distribution $\{p_{-j},p_{-j+1},\ldots,p_j\}$ while $j$ is presumed fixed.

\section{Wigner function reconstruction by filtered backprojection}
\label{sec:math}

The density matrix $\op{\rho}$ of a system of total angular momentum $j$ (assumed fixed here; this condition will be relaxed in \sref{sec:fluctj}) is usually expressed in one of the two forms
\numparts
\label{eq:rhommkq}
\begin{eqnarray}
	\label{eq:rhomm}
	\rho_{m m'} &=& \me{j m}{\op{\rho}}{j m'} = \sum_{k=0}^{2j} \sum_{q=-k}^k \rho_{k q} t_{k q}^{j m m'}\\
	\label{eq:rhokq}
	\rho_{k q} &=& \sum_{m=-j}^j \sum_{m'=-j}^j \rho_{m m'} t_{k q}^{j m m'},
\end{eqnarray}
\endnumparts
with the transformation coefficients (in the following simply termed Clebsch--Gordan coefficients)~\cite{Dowling1994}
\begin{equation}
	\label{eq:tkqjmm}
	t_{k q}^{j m m'} 
	= (-1)^{j-m-q}\scp{j,m;j,-m'}{k,q},
\end{equation}
nonzero only if $q=m-m'$.
Both forms contain the same information and are completely interchangeable. While form~\eref{eq:rhomm} is more common, form~\eref{eq:rhokq} allows expressing the Wigner function on the Bloch sphere~\eref{eq:Wsph}. Since our goal is the reconstruction of the Wigner function from experimental data, we focus on form~\eref{eq:rhokq}, in particular its low-$k$ components. 

In order to determine the unknown quantum-mechanical state of a system of total spin $j$, it is necessary that many instances of this state can be generated experimentally~\cite{QuantumStateEstimation}, on which destructive measurements are performed. Further, projective Stern--Gerlach measurements must be performed along many different quantization axis orientations $(\vartheta_n,\varphi_n)$. For the correctness of the following reconstruction method it is crucial that these measured quantization axes are distributed as evenly as possible over the hemisphere of orientations. Since this requirement may be difficult to fulfill experimentally, we assign weights $c_n$ to the individual measurements in order for the weighted measurement density to approximate a homogeneous distribution of quantization axes as best possible. Notice that these weights are independent of the outcomes $m_n$ of the Stern--Gerlach measurements. In the ideal case of homogeneously distributed quantization axis orientations (for example through the vertices of a geodesic hemisphere), all these weights are chosen equal and the data are used most efficiently.

In this way, the results from $M$ single Stern--Gerlach measurements along various quantization axes orientations are assembled into a data set of tuples $(\vartheta_n,\varphi_n,c_n,m_n)$ with $n=1\ldots M$ and $\sum_{n=1}^M c_n=1$. Our filtered backprojection algorithm for reconstructing the Wigner function coefficients is then given by
\begin{equation}
	\label{eq:fbp}
	\rho_{k q}\se{(fbp)} = (2k+1)\sum_{n=1}^M c_n D_{q 0}^{k}(\varphi_n,\vartheta_n,0) t_{k 0}^{j m_n m_n},
\end{equation}
with $D_{m' m}^{j}(\alpha,\beta,\gamma)=\me{j m'}{e^{-\rmi\alpha\op{J}_z} e^{-\rmi\beta\op{J}_y} e^{-\rmi\gamma\op{J}_z}}{j m}$ a Wigner rotation matrix~\cite{Zare}; in particular $D_{q 0}^k(\varphi,\vartheta,0) = \sqrt{\frac{4\pi}{2k+1}}Y_{k q}^*(\vartheta,\varphi)$. This is formally equivalent to the filtered backprojection algorithm used for planar inverse Radon transforms~\cite{KakSlaney}, with the factor $2k+1$ representing the ``filter'', and the summand representing the backprojection. Our algorithm has all of the typical properties of planar inverse Radon transforms by filtered backprojection: no data binning is required, and there are no \emph{ad hoc} parameters to be chosen or optimized. Further, as the backprojection algorithm is a direct sum and does not include an inversion step (such as a straight inversion of the Radon transform would require), the impact of experimental noise is bounded in the result. It is this last property which makes backprojection algorithms fast and reliable in practical applications such as X-ray computed tomography~\cite{KakSlaney}.

Our specific backprojection~\eref{eq:fbp} can be interpreted in an intuitive way. The measured values of $m_n$ in the coordinate frame attached to the quantization axis $(\vartheta_n,\varphi_n)$ are distributed according to the diagonal elements $\rho_{m_n m_n}$ and are converted from $j$-space into $k$-space via the Clebsch--Gordan coefficients $t_{k q'}^{j m_n m_n}$ with $q'=0$ (see \sref{app:ClebschGordan} for a numerical procedure). They are then rotated into the lab frame through the rotation matrices $D_{q q'}^k(\varphi_n,\vartheta_n,\chi_n)$ with the value of $\chi_n$ irrelevant (set to zero) since $q'=0$.

In the following, we demonstrate that this algorithm~\eref{eq:fbp} works in the limit of infinite data. If \emph{all} quantization axis orientations have been used with equal frequency, and infinitely many measurements have been performed along each quantization axis, the sum over measurements $\sum_{n=1}^M c_n$ can be replaced by a normalized integral $\frac{1}{2\pi}\int_0^{\pi/2}\sin\vartheta\rmd\vartheta\int_0^{2\pi}\rmd\varphi$ over the hemisphere of axis orientations (by symmetry the other hemisphere yields an identical result) and a sum over the measurement outcomes $m$,
\begin{equation}
	\fl
	\rho_{k q}\se{(fbp)} = \frac{2k+1}{2\pi}\int_0^{\pi/2}\sin\vartheta\rmd\vartheta\int_0^{2\pi}\rmd\varphi \sum_{m=-j}^j p_m(\vartheta,\varphi)
	D_{q 0}^{k} (\varphi,\vartheta,0) t_{k 0}^{j m m},
\end{equation}
where the Stern--Gerlach probability distribution along a quantization axis $(\vartheta,\varphi)$ is given by the diagonal elements $\rho_{m m}$ of~\eref{eq:rhomm} in the rotated frame,
\begin{equation}
	\label{eq:projprob}
	p_m(\vartheta,\varphi) = \sum_{k=0}^{2j} \sum_{q=-k}^k [D_{q 0}^{k}(\varphi,\vartheta,0)]^*\rho_{k q} t_{k 0}^{j m m}.
\end{equation}
Using the orthogonality relations of Clebsch--Gordan coefficients,
\begin{equation}
	\label{eq:tortho}
	\sum_{m=-j}^j t_{k 0}^{j m m}t_{k' 0}^{j m m} = \delta_{k k'},
\end{equation}
and spherical harmonics,
\begin{equation}
	\int_0^{\pi/2}\sin\vartheta\rmd\vartheta\int_0^{2\pi}\rmd\varphi [D_{q' 0}^k(\varphi,\vartheta,0)]^*D_{q 0}^k(\varphi,\vartheta,0) = \frac{2\pi}{2k+1} \delta_{q q'},
\end{equation}
it is easy to show that indeed $\rho_{k q}\se{(fbp)} =\rho_{k q}$, proving the validity of the reconstruction method in the limit of infinitely many homogeneously distributed Stern--Gerlach experiments.

In the more experimentally relevant case of a finite data set, the literature on the two-dimensional inverse Radon transform by filtered backprojection~\cite{KakSlaney} indicates that excellent results can still be recovered, albeit with aliasing artifacts present to some degree. As a rough estimate, if Stern--Gerlach experiments are performed only along certain quantization axes spaced by an average angle $\Delta\eta$, then the reconstructed partial waves of the Wigner function become unreliable for $k\gtrsim k\si{max}=\pi/\Delta\eta$. Further, if the number $M$ of measurements is much less than the number of degrees of freedom $(k\si{max}+1)^2$, then the reconstructed coefficients $\rho_{k q}$ will be dominated by noise, in particular at large $k$. Both of these effects are mitigated in \sref{sec:smooth} for the present reconstruction scheme.

\subsection{Accounting for fluctuations in the total angular momentum $j$}
\label{sec:fluctj}

We recall that for systems composed of many spin-$1/2$ components, such as two-component Bose--Einstein condensates, the total angular momentum $j=(N_{\uparrow}+N_{\downarrow})/2$ often varies between single Stern--Gerlach measurements, as each such measurement requires the preparation of a new condensate. Instead of constructing a separate Wigner function for each occurring value of $j$, we notice that for $k\ll2j$ the Clebsch--Gordan coefficients $t_{k 0}^{j m m}$ depend smoothly on the total angular momentum $j$. This allows us to reconstruct the low-resolution part of the Wigner function even if $j$ varies slightly between single Stern--Gerlach measurements. To this end we include the measured values of $j$ in the data tuples, extending them to $(\vartheta_n,\varphi_n,c_n,j_n,m_n)$; the filtered backprojection formula is modified to
\begin{equation}
	\label{eq:fbpj}
	\rho_{k q}\se{(fbp)} = (2k+1)\sum_{n=1}^M c_n D_{q 0}^{k}(\varphi_n,\vartheta_n,0) t_{k 0}^{j_n m_n m_n}.
\end{equation}
Again we refer to \ref{app:ClebschGordan} for a numerical method to evaluate this expression.

The same smoothness of the Clebsch--Gordan coefficients at low $k$ is used in \sref{sec:smooth} to treat measurement uncertainties in both $j_n$ and $m_n$ in a perturbative manner in~\eref{eq:fbpj}. This is fundamentally different from a direct tomographic reconstruction of the Dicke matrix elements $\rho_{m m'}$, where such uncertainties introduce large but correlated errors throughout the density matrix and make such a perturbative treatment impossible.

\subsection{Measurement uncertainties and high-$k$ damping}
\label{sec:smooth}

It is natural to assume that $M$ uncorrelated experimental measurements can only serve to reconstruct $M$ coefficients $\rho_{k q}$, suggesting an upper limit $k\si{max}\approx \sqrt{M}$ (assuming again a homogeneous distribution of quantization axis orientations). For larger values of $k$ the angular power spectrum~\cite{Hinshaw2003b}
\begin{equation}
	\label{eq:spectrum}
	C_k\se{(fbp)} = \frac{1}{2k+1} \sum_{q=-k}^k |\rho_{k q}\se{(fbp)}|^2
\end{equation}
tends to acquire large fluctuations because of insufficient experimental data (see \fref{fig:spectrum} for an example). However, simply cutting the reconstruction off at $k\si{max}$ is unsatisfactory because it disregards that some useful information is still present in these high-$k$ partial waves. A more natural cutoff is introduced through the $k$-dependent sensitivity to experimental uncertainties. Assuming experimental variances of $\avg{N_{\uparrow}^2}-\avg{N_{\uparrow}}^2=\avg{N_{\downarrow}^2}-\avg{N_{\downarrow}}^2=\sigma_N^2$, we find that the uncertainties of $\avg{j^2}-\avg{j}^2=\avg{m^2}-\avg{m}^2=\sigma_N^2/2$ (with no covariance, $\avg{j m}=\avg{j}\avg{m}$) yield a leading order damping of the Clebsch--Gordan coefficients
\begin{equation}
	\label{eq:tdamp}
	\avg{t_{k 0}^{j m m}} \approx t_{k 0}^{j m m} \exp\left[-\frac{\sigma_N^2}{2j(2j-1)}k(k+1)\right].
\end{equation}
The rotation matrix elements are damped similarly: if the pointing direction of the quantization axis $\Omega=(\vartheta,\varphi)$ has an uncertainty of $\sigma_{\Omega}\ll1$ \{in terms of the expectation value of the angle $\eta_{\Omega \Omega'}$ between the ideal axis orientation $\Omega$ and its true experimental value $\Omega'$ we define $\sigma_{\Omega}^2 = \avg{\sin^2\eta_{\Omega \Omega'}}=\avg{1-[\cos\vartheta\cos\vartheta'+\sin\vartheta\sin\vartheta'\cos(\varphi-\varphi')]^2}$\}, then for large $k$ we find the rotation matrix elements to be damped as
\begin{equation}
	\label{eq:Ddamp}
	\avg{D_{q 0}^{k}(\varphi,\vartheta,0)} \approx D_{q 0}^{k}(\varphi,\vartheta,0) \exp\left[-\frac{\sigma_{\Omega}^2}{4}k(k+1)\right].
\end{equation}
If $\sigma_N$ and $\sigma_{\Omega}$ are equal for all measurements, the linearity of~\eref{eq:fbpj} yields a simple smoothing $\rho_{k q}\mapsto\rho_{k q}e^{-\alpha k(k+1)}$ with $\alpha=\frac{\sigma_N^2}{2j(2j-1)}+\frac{\sigma_{\Omega}^2}{4}$. In this way, these two damping formulas~(\ref{eq:tdamp},\ref{eq:Ddamp}) cut off the reconstruction at large $k$ in a natural and smooth way.

\subsection{Assembling the Wigner function}
\label{sec:dW}

\begin{figure}
	\begin{centering}
	\includegraphics[width=0.7\textwidth]{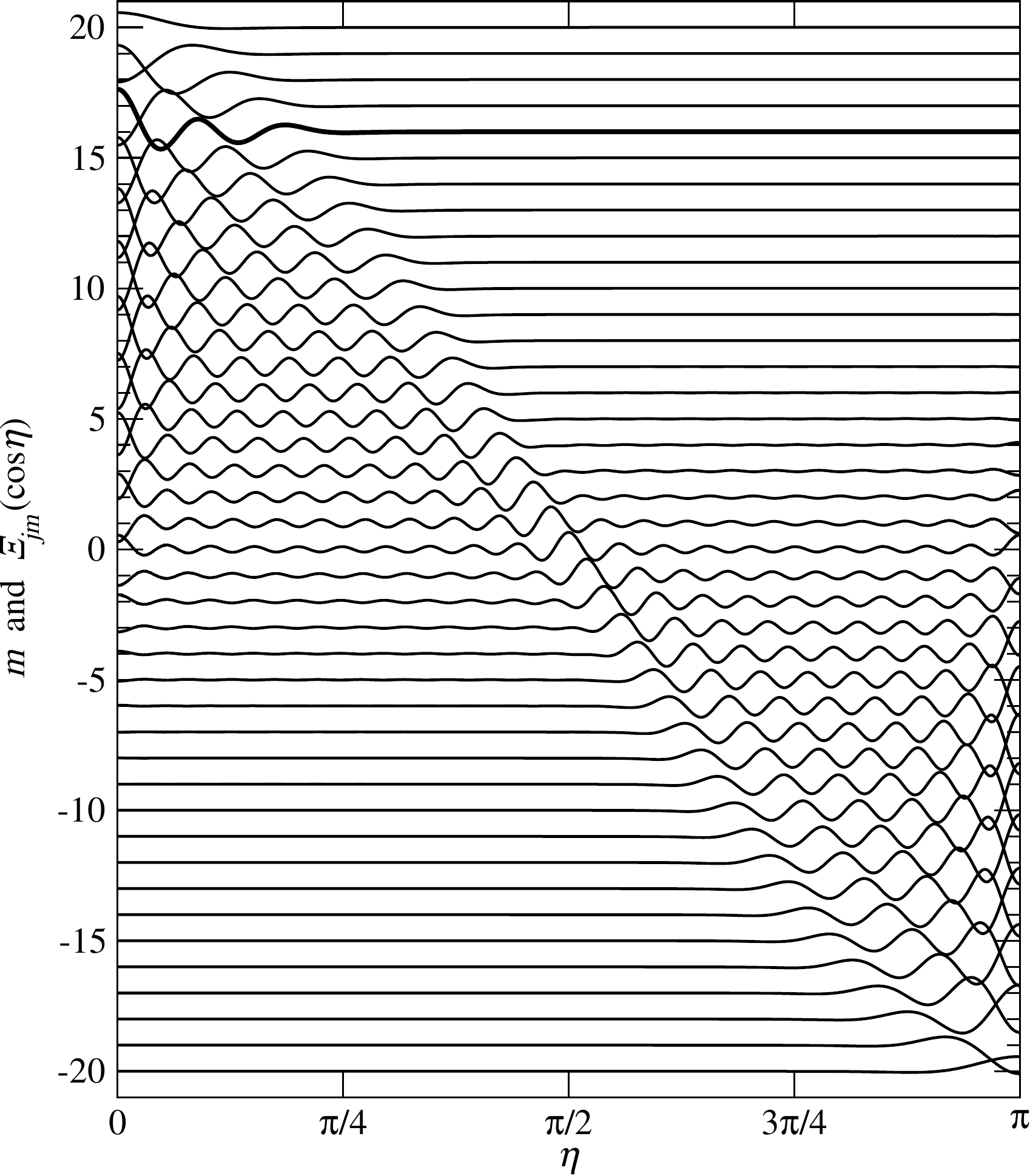}
	\caption{Contributions $\Xi_{j m}(\cos\eta)$ to the Wigner function~\eref{eq:dW} for $j=20$ and $m=-20\ldots+20$. All curves have been divided by 100 and offset vertically by $m$. The bold curve for $m=+16$ is used in \fref{fig:dW}. Notice that the $m=\pm j$ contributions have lower spatial resolution ($\Delta\eta\sim1/\sqrt{j}$) than those with $m\approx 0$ ($\Delta\eta\sim1/j$); see \sref{sec:planar}.}
	\label{fig:dWall}
	\end{centering}
\end{figure}

\begin{figure}
	\begin{centering}
	\includegraphics[width=0.4\textwidth]{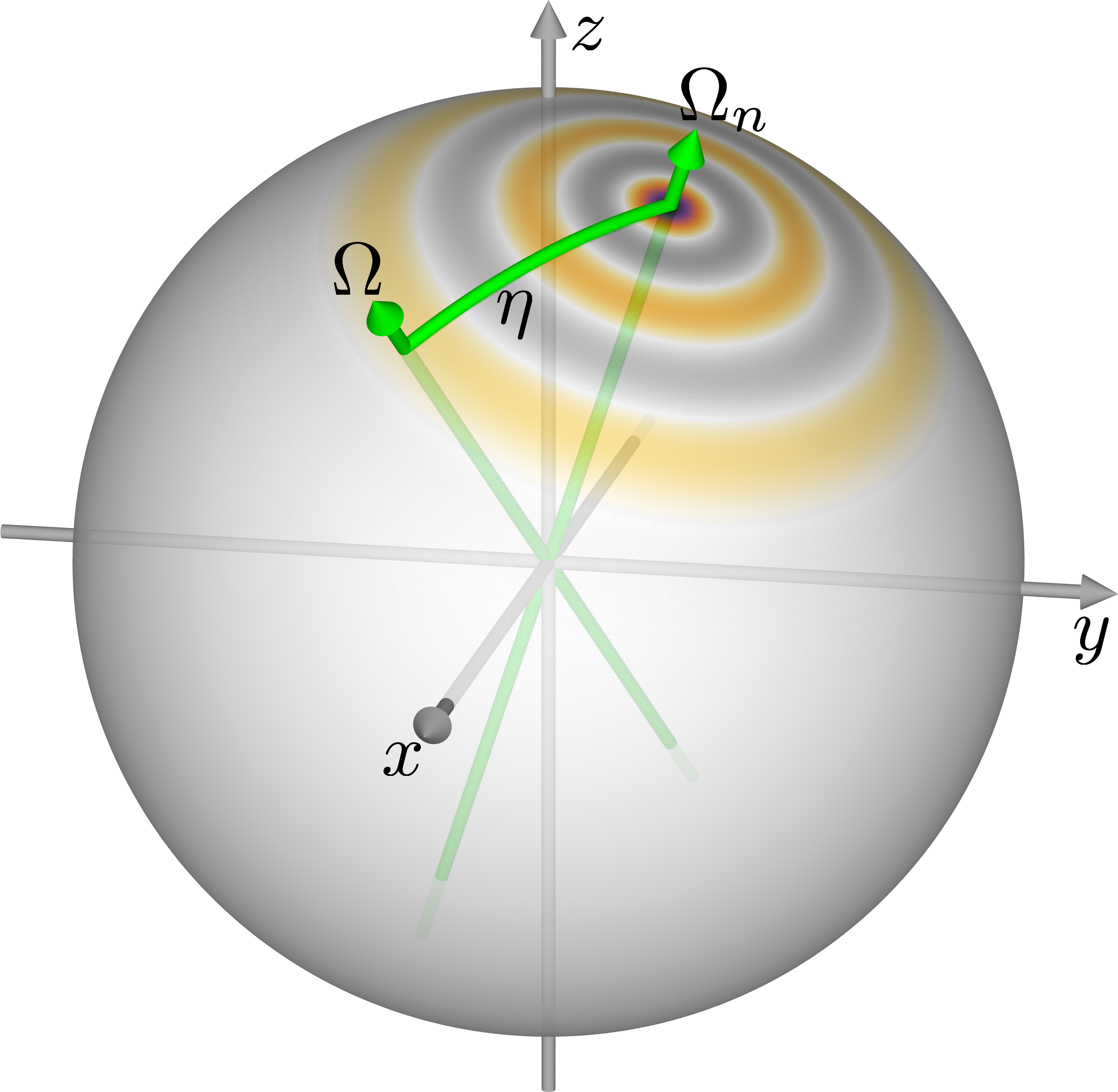}
	\caption{Contribution to the Wigner function~\eref{eq:dW} for $j_n=20$ and $m_n=16$ (see \fref{fig:dWall}); colors as in \fref{fig:Wignerfunction} but scaled to the maximum value of $+163$. The contribution $\Xi_{20,16}(\cos\eta)$ depends only on the angle $\eta$ between the quantization axis $\Omega_n$ and the direction $\Omega$ in which the Wigner function is measured.}
	\label{fig:dW}
	\end{centering}
\end{figure}

Inserting the resulting coefficients~\eref{eq:fbpj} into the form of the Wigner function~\eref{eq:Wsph} we find the tomographically reconstructed Wigner function
\begin{eqnarray}
	\fl
	W\se{(fbp)}(\vartheta,\varphi) &=& \sum_{n=1}^M c_n \left[ \sum_{k=0}^{2j} \sum_{q=-k}^k (2k+1) D_{q 0}^{k}(\varphi_n,\vartheta_n,0) Y_{k q}(\vartheta,\varphi) t_{k 0}^{j_n m_n m_n} \right]\nonumber\\
	 &=& \sum_{n=1}^M c_n \Xi_{j_n,m_n}[\cos\vartheta\cos\vartheta_n+\sin\vartheta\sin\vartheta_n\cos(\varphi-\varphi_n)],
\end{eqnarray}
where the contributions can be simplified to
\begin{equation}
	\label{eq:dW}
	\Xi_{j m}(x) = \frac{1}{\sqrt{4\pi}} \sum_{k=0}^{2j} (2k+1)^{3/2}\, t_{k 0}^{j m m} P_k(x).
\end{equation}
As is to be expected in spherical symmetry, the contribution of an individual Stern--Gerlach measurement (see \fref{fig:dWall}) depends only on the relative angle $\cos\eta_{\Omega \Omega_n}=\cos\vartheta\cos\vartheta_n+\sin\vartheta\sin\vartheta_n\cos(\varphi-\varphi_n)$ between the quantization axis orientation $\Omega_n=(\vartheta_n,\varphi_n)$ of the measurement and the point $\Omega=(\vartheta,\varphi)$ on the Bloch sphere (\fref{fig:dW}).
Similarly to technical implementations of the planar inverse Radon transform~\cite{KakSlaney}, the Wigner function is thus assembled from additive contributions due to the individual Stern--Gerlach measurements (see \fref{fig:dW} for an example). The constructive or destructive interference of these contributions is what yields the reconstructed Wigner function (see \fref{fig:Wignerfunction}). The spatial resolutions of the $\Xi_{j m}(\cos\eta)$ ultimately determine the spatial resolution of the reconstructed Wigner function: if the Wigner function is composed predominantly of contributions with $m_n\approx \pm j_n$ its angular resolution is limited by that of a coherent state, $\Delta\eta\gtrsim1/\sqrt{\avg{j}}$; if on the other hand the majority of contributions has $m_n\approx 0$ the resolution can be significantly higher, $\Delta\eta\gtrsim1/\avg{j}$. We make use of this observation in sections~\ref{sec:planar} and~\ref{sec:data}, where a spin-squeezed state is reconstructed and the increased spatial resolution is critically important.

\subsection{Positivity of the density matrix}
\label{sec:positivity}

It is well known that only positive semi-definite density matrices represent valid quantum-mechanical states of a system~\cite{QuantumStateEstimation}. Unfortunately, the filtered backprojection method~\eref{eq:fbpj} does not assure that the reconstructed $\op{\rho}$ is positive semi-definite when used with a finite and noisy data set. For the purpose of displaying the Wigner function graphically, this is of no concern (see \fref{fig:Wignerfunction}); however, when the tomographically reconstructed coefficients $\rho_{k q}\se{(fbp)}$ are used in quantitative calculations (see \sref{sec:data}) positivity can be crucial. This is a similar problem as the requirement for a positive absorption density in medical computed tomography (CT) imaging~\cite{KakSlaney}. It is also present in many quantum-state reconstruction schemes, and has been discussed extensively in the quantum tomography literature~\cite{QuantumStateEstimation}.

We do not offer a solution for assuring the positivity of the reconstructed density matrix. Here we merely point out that in other reconstruction schemes, such as maximum-likelihood estimates~\cite{Banaszek2000}, the \emph{ansatz} $\op{\rho}=\op{T}\dagg\op{T}$ forces the density matrix $\op{\rho}$ to be positive semi-definite; but a direct tomographic reconstruction of $\op{T}$ similar to~\eref{eq:fbpj} is currently lacking.

\section{Quantization axes lying in a single plane}
\label{sec:planar}

When the spin-$j$ system's quantum-mechanical state is fairly localized on the Bloch sphere, not every choice of quantization axis orientation has the same potential for extracting information about the state. When the axis is close to \emph{parallel} to the state, most Stern--Gerlach measurements will yield $|m|\approx j$, with a limited angular resolution $\sim1/\sqrt{j}$ given by the size of a coherent state on the Bloch sphere~\cite{Agarwal1998}. If the axis is close to \emph{perpendicular} to the state, on the other hand, the distribution of measured values $m$ represents the structure of the state's Wigner function much more accurately, with an angular resolution $\sim1/j$. This difference in scaling of the angular resolution, visible in \fref{fig:dWall}, suggests that for large $j$ it may be advantageous to focus on performing Stern--Gerlach measurements with quantization axes in a plane perpendicular to the quantum state, instead of covering the entire hemisphere of axis orientations. As a consequence much fewer measurements are needed, and we can get much more rapid convergence of the reconstruction in practice. But it is not \emph{a priori} clear that this restriction of the quantization axes to a single plane has the potential for reconstructing the full quantum-mechanical state of the system.

As it turns out, a modification to the ``filter'' function in~\eref{eq:fbpj} results in a full reconstruction of the mirror-symmetric part of the Wigner function. Defining the coordinate system such that the state is localized near the $+z$ axis and all quantization axes lie in the $x y$ plane, the \emph{in-plane} filtered backprojection formula is
\begin{equation}
	\label{eq:fbpjP}
	\rho_{k q}\se{(fbp,P)} = (\case{k-q+1}{2})_{\frac12}(\case{k+q+1}{2})_{\frac12} \pi \sum_{n=1}^M c_n D_{q 0}^{k}(\varphi_n,\frac{\pi}{2},0) t_{k 0}^{j_n m_n m_n},
\end{equation}
where $(a)_n=\Gamma(a+n)/\Gamma(a)$ is a Pochhammer symbol, and $(a)_{\frac12}\approx\sqrt{a}-1/(8\sqrt{a})$.

We again prove this reconstruction in the infinite-data limit. In the case of a homogeneous distribution of all azimuthal axis orientation angles $\varphi$ we use the relationship
\begin{eqnarray}
	\fl
	\frac{1}{\pi} \int_0^{\pi}\rmd\varphi [D_{q' 0}^k(\varphi,\frac{\pi}{2},0)]^*D_{q 0}^k(\varphi,\frac{\pi}{2},0)\nonumber\\
	= \cases{\frac{\delta_{q q'}}{(\frac{k-q+1}{2})_{\frac12}(\frac{k+q+1}{2})_{\frac12}\pi} & if $k+q$ even\\
	0 & if $k+q$ odd,}
\end{eqnarray}
which remains true in the experimentally more relevant case of a finite number $A$ of equally-spaced axis orientations (replacing $\frac{1}{\pi}\int_0^\pi\rmd\varphi \mapsto \frac{1}{A}\sum_{a=0}^{A-1}$ with $\varphi=a\pi/A$) as long as $k<A$. Together with~\eref{eq:projprob} and~\eref{eq:tortho} we thus find that
\begin{eqnarray}
	\fl
	\rho_{k q}\se{(fbp,P)} =
	(\case{k-q+1}{2})_{\frac12}(\case{k+q+1}{2})_{\frac12} \int_0^{\pi}\rmd\varphi\, p_m(\frac{\pi}{2},\varphi) D_{q 0}^{k}(\varphi,\frac{\pi}{2},0) t_{k 0}^{j_n m_n m_n}\nonumber\\
	= \cases{\rho_{k q}&if $k+q$ even\\0&if $k+q$ odd.}
\end{eqnarray}
Thus in the infinite-data limit such an in-plane reconstruction exactly determines the coefficients $\rho_{k q}$ for which $k+q$ is even, while giving no information on the coefficients for which $k+q$ is odd. Since the parity of $k+q$ is the $z\leftrightarrow-z$ reflection parity of the spherical harmonics $Y_{k q}(\vartheta,\varphi)$, the in-plane formula~\eref{eq:fbpjP} reconstructs the positive-parity component $W^+(\vartheta,\varphi)$ of the Wigner function $W(\vartheta,\varphi)=W^+(\vartheta,\varphi)+W^-(\vartheta,\varphi)$, with $W^{\pm}(\pi-\vartheta,\varphi)=\pm W^{\pm}(\vartheta,\varphi)$. If we know from other measurements that the state is fully localized on the ``northern'' Bloch hemisphere ($z>0$), then the correct Wigner function is
\begin{equation}
	W(\vartheta,\varphi) = \cases{2W^+(\vartheta,\varphi) & if $0\le\vartheta<\frac{\pi}{2}$\\0& if $\frac{\pi}{2}<\vartheta\le\pi$,}
\end{equation}
which has the decomposition
\begin{equation}
	\fl
	\rho_{k q}\se{(fbp,P,N)} = \int_0^{\pi}\sin\vartheta\rmd\vartheta\int_0^{2\pi}\rmd\varphi\, Y_{k q}^*(\vartheta,\varphi)W(\vartheta,\varphi)
	=\sum_{k'=0}^{2j} \Upsilon_{k k'}^q \rho_{k' q}\se{(fbp,P)}
\end{equation}
in terms of the overlap integrals $\Upsilon_{k k'}^q$ given in \ref{app:overlap}. We conclude that the data acquired by Stern--Gerlach measurements with quantization axes lying solely within a plane are sufficient for an exact reconstruction of the Wigner function.

\subsection{Measurement uncertainties and high-$k$ damping}
\label{sec:smoothP}

Measurement uncertainties can be introduced in~\eref{eq:fbpjP} in the same way as in \sref{sec:smooth}. However, in an in-plane measurement series we can additionally separate out the azimuthal axis orientation uncertainty: since the rotation matrix elements $D_{q 0}^k(\varphi,\vartheta,0)$ are proportional to $e^{-\rmi q\varphi}$, a variance $\avg{\varphi^2}-\avg{\varphi}^2=\sigma_{\varphi}^2$ leads to a damping
\begin{equation}
	\label{eq:DdampP}
	\avg{D_{q 0}^k(\varphi,\frac{\pi}{2},0)} = D_{q 0}^k(\varphi,\frac{\pi}{2},0) \exp(-\frac12 q^2\sigma_{\varphi}^2).
\end{equation}

\section{Demonstration with experimental data}
\label{sec:data}

In this section we reconstruct the Wigner function from a data set describing ensembles of $N= 1250(45)$ atoms acquired in our group~\cite{Riedel2010}. In contrast to~\cite{Riedel2010} we rotate the coordinate system such that all quantization axes lie in the $x y$ plane and the state is localized around the $+z$ axis; in this way the procedure of \sref{sec:planar} can be employed directly. The data set consists of three experimental runs spanning different ranges of $\varphi$ with different angular resolutions, owing to the fact that the need for homogeneity in $\varphi$ for the filtered backprojection algorithm~\eref{eq:fbpjP} was not known at the time of data acquisition. We use weights $c_n$ adjusted such that the weighted density of Stern--Gerlach measurements is as close to homogeneous as possible over the range $\varphi=0\ldots\pi$ of azimuthal quantization axis orientations. As discussed in \sref{sec:planar} the planar arrangement of quantization axis orientations leads to a Wigner function which is peaked along both the $+z$ and $-z$ directions, featuring two identical copies of the quantum state. An additional Ramsey experiment~\cite{Riedel2010} was used to experimentally determine the correct location of the state on the northern ($z>0$) Bloch hemisphere.

\begin{figure}
	\begin{centering}
	\includegraphics[width=0.7\textwidth]{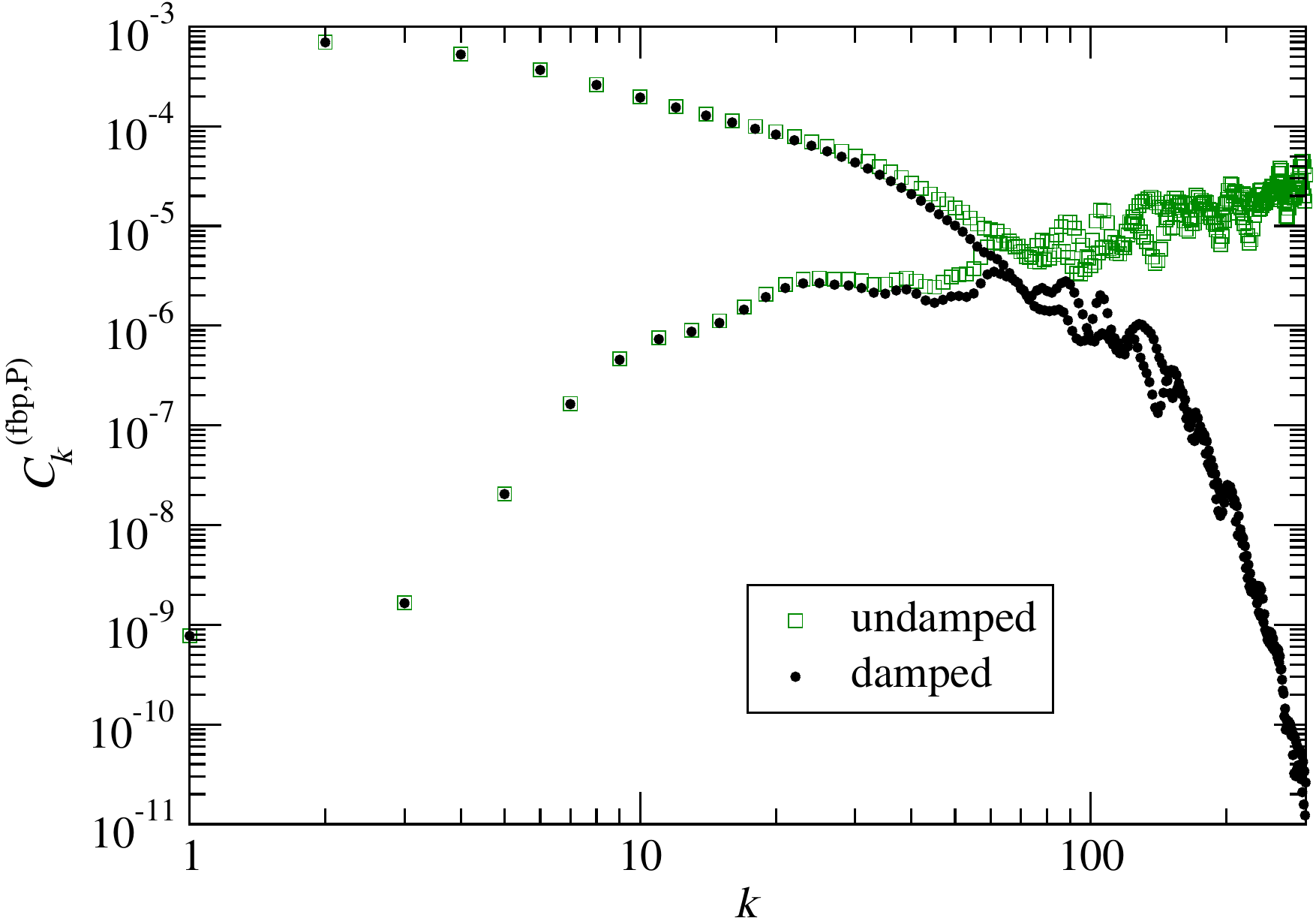}
	\caption{Angular power spectrum~\eref{eq:spectrum} of the reconstructed Wigner function of \fref{fig:Wignerfunction}. Without damping ({\color{bulletcolor}\tiny $\boxempty$}) the power in modes $k\gtrsim70$ is too large and dominated by noise and aliasing effects; experimental uncertainties damp the angular power at large $k$ in a natural way ($\bullet$, see sections~\ref{sec:smooth} and~\ref{sec:smoothP}). Odd-$k$ modes contain less power than even-$k$ modes because of the approximate point symmetry of the Wigner function (see \fref{fig:Wignerfunction}).}
	\label{fig:spectrum}
	\end{centering}
\end{figure}

High-$k$ damping (\sref{sec:smoothP}) is achieved with an experimental uncertainty of $\sigma_N\approx11$ atoms~\eref{eq:tdamp} and with an experimental error model dominated by phase noise: $\sigma_{\varphi}\approx \sigma\si{ph}^2\sin(|\varphi|)/\sqrt{2}$ in~\eref{eq:DdampP}, with phase noise amplitude $\sigma\si{ph}=8.2^{\circ}$~\cite{Riedel2010}. In \fref{fig:spectrum} the effect of this damping is shown to be crucial for partial waves $k\gtrsim70$.

\begin{figure}
	\centering
	\includegraphics[width=0.7\textwidth]{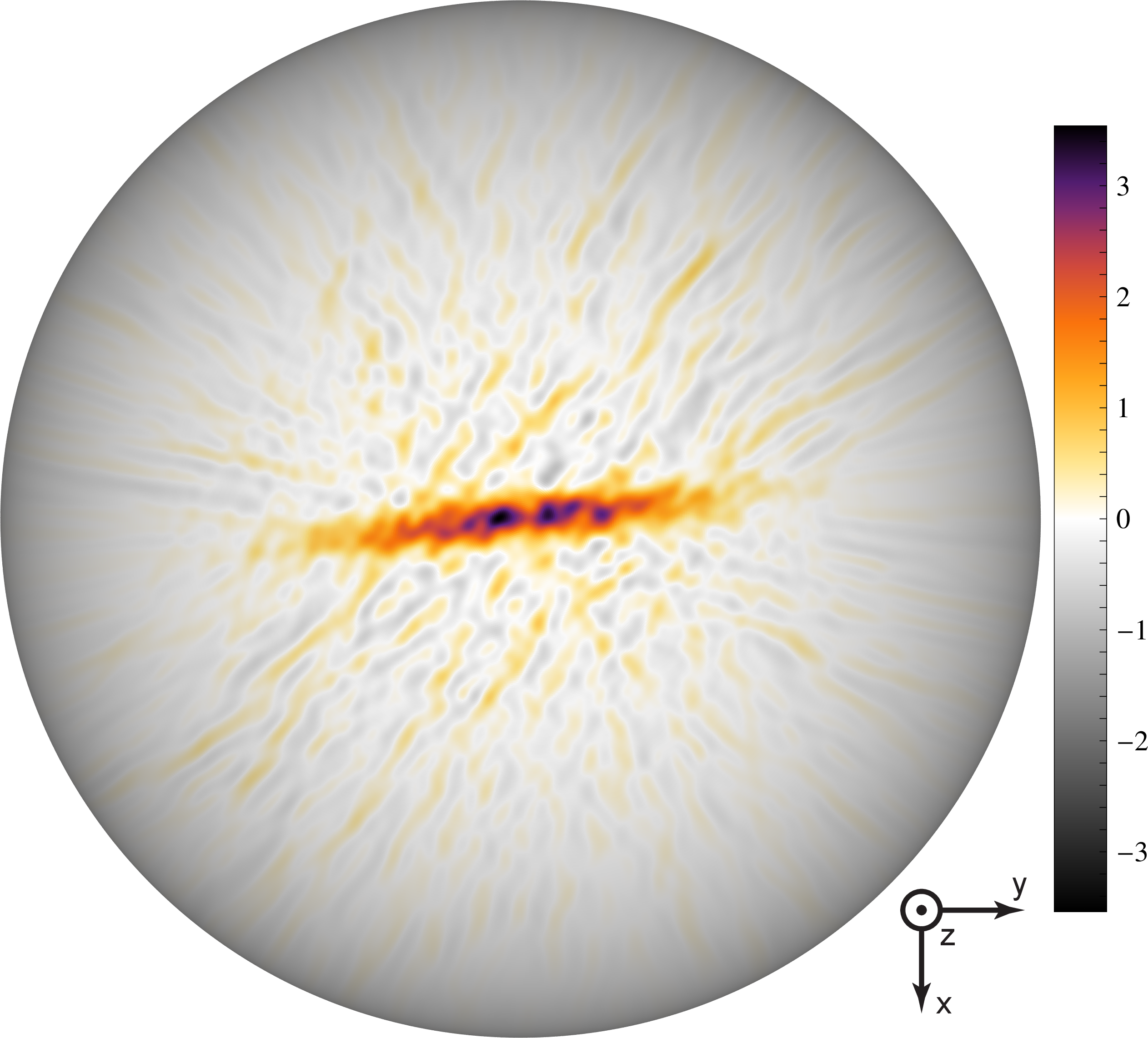}
	\caption{Reconstructed Wigner function using the data set described in \sref{sec:data}. The Wigner function takes values from $-0.93$ to $+3.54$. The coordinate system is rotated from~\cite{Riedel2010} (see text).}
	\label{fig:Wignerfunction}
\end{figure}

The resulting reconstructed Wigner function is shown in \fref{fig:Wignerfunction}. The high-frequency artifacts in the Wigner function far from the central peak are due to incomplete destructive interference of the contributions from the individual Stern--Gerlach measurements (see \sref{sec:dW}). We expect that a more complete data set, including more quantization axis orientations, will lead to a smoother Wigner function at large angles $\vartheta$.

\subsection{Spin-squeezing measurement}

\begin{figure}
	\centering
	\includegraphics[width=0.7\textwidth]{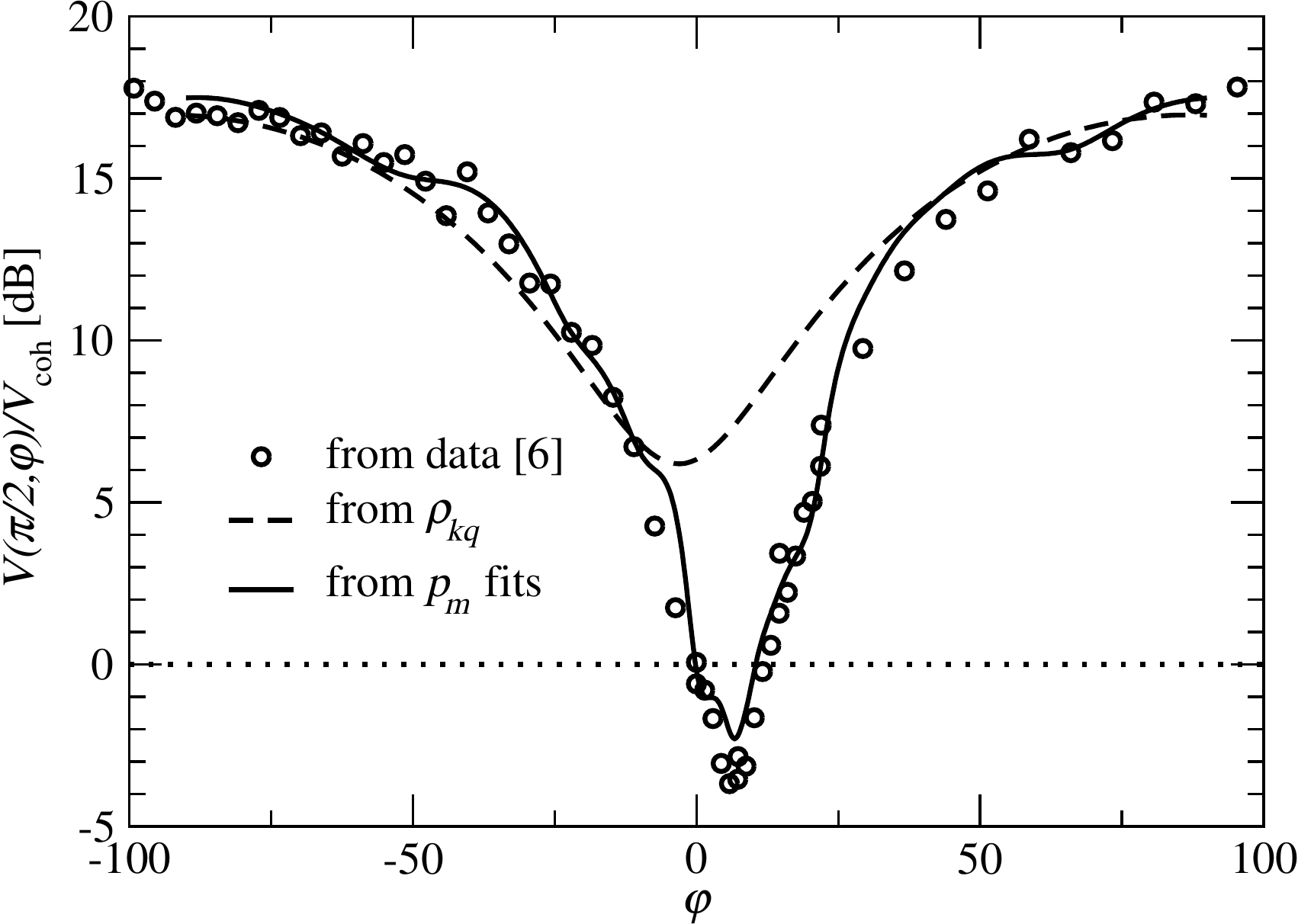}
	\caption{Normalized variance $V=\avg{m^2}-\avg{m}^2$ as a function of azimuthal quantization axis orientation $\varphi$. Open circles show variances calculated directly from Stern--Gerlach experiments along a given quantization axis~\cite{Riedel2010}. The dashed line was calculated directly from the coefficients $\rho_{k q}\se{(fbp,P)}$ through~\eref{eq:expect}. The solid line shows the results of Gaussian fits (\fref{fig:Gfits}) to the probability distributions $p_m(\frac{\pi}{2},\varphi)$ given in~\eref{eq:projprob}. As in~\cite{Riedel2010} we first subtract the experimental noise ($\sigma_N^2/2$ with $\sigma_N=11$) from the calculated variances, and then divide by the variance of a coherent state, $V\si{coh}=\avg{j}/2$ with $\avg{j}=630$ [see~\eref{eq:cohvar}]. A spin-squeezed state is characterized by negative values (in dB).}
	\label{fig:expect}
\end{figure}

\begin{figure}
	\centering
	\includegraphics[width=0.7\textwidth]{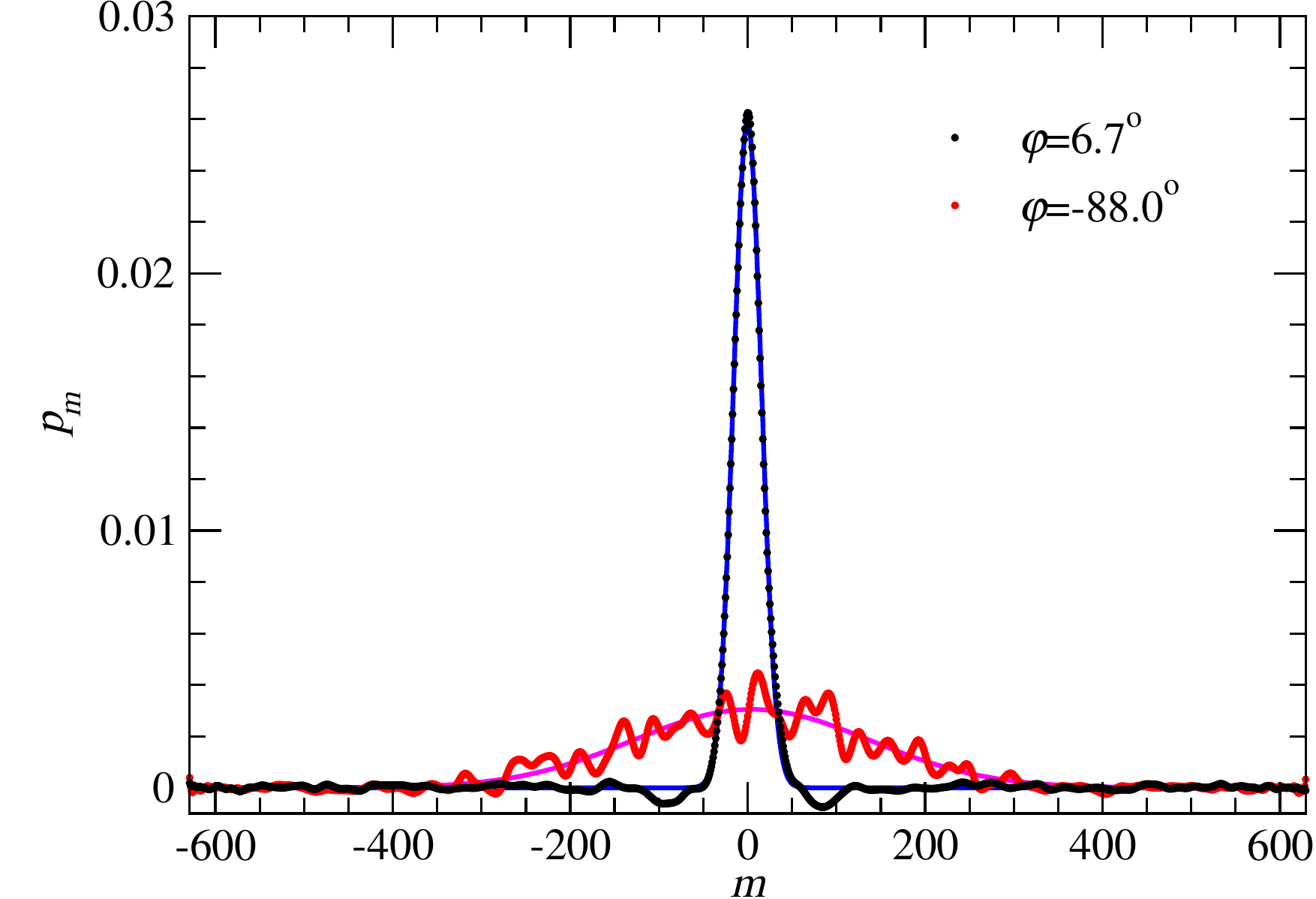}
	\caption{Probability distributions $p_m$ of the projection quantum number $m$ along the minimum-variance axis ($\varphi\si{s}=6.7^{\circ}$) and the maximum-variance axis ($\varphi=-88.0^{\circ}$) of \fref{fig:expect}, assuming $j=629$ [found from $\rho_{0 0}=\avg{(2j+1)^{-1/2}}\approx0.02818$]. Negative values of $p_m$ indicate that the reconstructed density matrix (as shown in \fref{fig:Wignerfunction}) is not positive semi-definite and therefore does not strictly represent a physical state (see \sref{sec:positivity}). Gaussian fits used for \fref{fig:expect} are shown as continuous lines.}
	\label{fig:Gfits}
\end{figure}

We demonstrate the quantitative use of the reconstructed Wigner function by estimating the amount of spin squeezing in the system.
Given a set of reconstructed Wigner function coefficients $\rho_{k q}$ we can calculate the probability distribution for the angular momentum projection quantum number onto any quantization axis orientation $(\vartheta,\varphi)$ from~\eref{eq:projprob}.
In principle the variance $V(\vartheta,\varphi)=\avg{m^2}(\vartheta,\varphi)-[\avg{m}(\vartheta,\varphi)]^2$ measures the amount of spin noise obtained experimentally. The expectation values of small integer powers of the projection quantum number $m$ depend only on the low-$k$ components of the Wigner function, which are particularly insensitive to experimental noise~(\ref{eq:tdamp},\ref{eq:Ddamp},\ref{eq:DdampP}); in particular,
\begin{eqnarray}
	\label{eq:expect}
	\fl
	\avg{m}(\vartheta,\varphi) = \sum_{m=-j}^j m\, p_m(\vartheta,\varphi)= \sqrt{\frac{(2j)_3}{12}} \sum_{q=-1}^1 [D_{q 0}^{1}(\varphi,\vartheta,0)]^*\rho_{1 q}\nonumber\\
	\fl
	\avg{m^2}(\vartheta,\varphi) = \sum_{m=-j}^j m^2 p_m(\vartheta,\varphi) \nonumber\\
	= \frac{j(j+1)\sqrt{2j+1}}{3} \rho_{0 0}
	+\sqrt{\frac{(2j-1)_5}{180}}\sum_{q=-2}^2 [D_{q 0}^{2}(\varphi,\vartheta,0)]^*\rho_{2 q}.
\end{eqnarray}
In \fref{fig:expect} we plot the resulting variances for quantization axes in the $x y$ plane, and compare them to a coherent state centered on the $+z$ axis. In the presence of imaging noise the variance of such a coherent state is given by~\eref{eq:expect} with $\rho_{k q}\se{(coh)}=t_{k q}^{j j j} e^{-\frac{\sigma_N^2}{2j(2j-1)} k(k+1)}$,
\begin{equation}
	\fl
	\label{eq:cohvar}
	\avg{m^2}\se{(coh)}
	= \frac{j(j+1)}{3}-\frac{j(2j-1)}{6}e^{-\frac{6\sigma_N^2}{2j(2j-1)}}
	= \frac{j+\sigma_N^2}{2} + \mathcal{O}(\sigma_N^4/j^2).
\end{equation}
In \fref{fig:expect} the experimental variance and the coherent-state variance are compared \emph{without} imaging noise, \emph{i.e.}, the leading-order imaging noise contribution $\sigma_N^2/2$ is subtracted from the experimental variance before comparison with the variance of a noise-free coherent state.
We notice that the resulting variance of the reconstructed state (dashed line in \fref{fig:expect}) is much larger than what was determined directly from the variances of the Stern--Gerlach data sets along the different quantization axes (open circles). We believe that this is a result of the lack of positivity of the density matrix (see \sref{sec:positivity}), owing to the finite and noisy data set used for its tomographic reconstruction. In fact, the probability distributions in \fref{fig:Gfits} clearly show negative values, which strictly speaking render the reconstructed density matrix unphysical.

As an alternative extraction method we calculate the probability distribution $p_m(\vartheta,\varphi)$ along a given quantization axis through~\eref{eq:projprob} and fit it with a Gaussian curve (see \fref{fig:Gfits}); the variance of this fit then serves as an estimate of $V(\vartheta,\varphi)$. In such a fit the positivity of the $p_m$ is no longer a crucial ingredient. In \fref{fig:expect} we show that this produces results that are very close to the variances calculated directly from Stern--Gerlach experiments along the various quantization axes. The deviations close to the squeezing maximum ($\varphi\si{s}\approx6.7^{\circ}$) result from the fact that the reconstructed Wigner function contains contributions from \emph{all} measurements, and therefore the extracted variance along a given quantization axis may be contaminated. Nonetheless the reconstructed Wigner function delivers a very concise picture of the structure of the multiparticle state, even for a data set with a non-uniform distribution over quantization axis orientations and with fluctuating values of $j_n$. Further, with our method the variance $V(\vartheta,\varphi)$ can be calculated along \emph{any} quantization axis orientation.

In practice any proof of spin squeezing will not proceed through the reconstruction of the Wigner function followed by either a fit to the projection~\eref{eq:projprob} or a direct study of the projection noise~\eref{eq:expect}. Instead, once the direction of squeezing $\varphi\si{s}$ has been determined, a full Stern--Gerlach experiment will be performed along this axis in order to directly estimate the probability distribution $p_m(\varphi\si{s})$, as in~\cite{Riedel2010} and in \fref{fig:expect} (circles). In this way, problems associated with the positivity of the reconstruction (\sref{sec:positivity}) and with the influence of experimental data and noise from directions $\varphi\neq\varphi\si{s}$ are strictly eliminated.

In future experiments providing data for the present tomographic reconstruction method, we plan to perform Stern--Gerlach measurements along many more quantization axes, but with as little as a single measurement per axis. Further, we will pay attention to cover the entire range of quantization axes uniformly [either the entire equator for~\eref{eq:fbpjP} or the entire sphere for~\eref{eq:fbpj}] in each experimental run. In this way, we expect to need only minimal data preprocessing before reconstructing the Wigner function, and will be able to use the acquired data in the most efficient way by using equal weights $c_n=1/M$ for all data points. We also expect that for such an improved data set the variance of the simple estimate given by~\eref{eq:expect} will be closer to that of the quantum-mechanical state.

\section{Conclusions}

We have presented a simple method for a tomographic reconstruction of the Wigner function of a spin-$j$ system, applicable even to experimental settings where $j$ is large and fluctuates between measurements. While the general procedure~\eref{eq:fbpj} requires Stern--Gerlach type measurements spread uniformly over all possible quantization axis orientations, a more specialized and faster procedure~\eref{eq:fbpjP} determines the Wigner function using only a single plane of quantization axis orientations. We have shown that this latter procedure is capable of reconstructing the Wigner function of a spin-squeezed state from a recently published experimental data set~\cite{Riedel2010}.

\ack
We thank Jonathan Dowling and Wolfgang Schleich for helpful discussions, and Max Riedel for help with the interpretation of the experimental data. This research was supported by the Swiss National Science Foundation and by the European Community through the project AQUTE.

\appendix

\section{Numerically evaluating Clebsch--Gordan coefficients}
\label{app:ClebschGordan}

We have used a recursion relation~\cite{Schulten1975} to evaluate the Clebsch--Gordan coefficients $\tau_k^{j,m}=t_{k 0}^{j m m}=(-1)^{j-m}\scp{j,m;j,-m}{k,0}$ from~\eref{eq:tkqjmm}:
\begin{eqnarray}
	\fl
	\tau_k^{j,j} = \frac{\pi^{1/4}\sqrt{2k+1}}{2^{2j+1/2}}
	\sqrt{\frac{\binom{4j+1}{2j-k}}{(2j+1)_{\frac12}}}\nonumber\\
	\fl
	\tau_k^{j,j-1} = \left( 1-\frac{k(k+1)}{2j} \right) \tau_k^{j,j}\nonumber\\
	\fl
	\tau_k^{j,m} = \frac{2j(j+1)-2(m+1)^2-k(k+1)}{j(j+1)-m(m+1)}\, \tau_k^{j,m+1}
		-\frac{j(j+1)-(m+1)(m+2)}{j(j+1)-m(m+1)}\, \tau_k^{j,m+2}\nonumber\\
	\fl
	\tau_k^{j,-m} = (-1)^k \tau_k^{j,m}.
\end{eqnarray}
This procedure is numerically stable even at very large values of $j$ and $k$.

\section{Hemispherical overlap integrals of spherical harmonics}
\label{app:overlap}

The hemispherical overlap integrals of the spherical harmonics are~\cite{Ashour1964}
\begin{eqnarray}
	\fl
	\Upsilon_{k k'}^q = 2\int_0^{\pi/2}\sin\vartheta\rmd\vartheta\int_0^{2\pi}\rmd\varphi\, Y_{k q}^*(\vartheta,\varphi) Y_{k' q}(\vartheta,\varphi)\nonumber\\
	= \cases{
		1\ \mathrm{if}\ k=k'\\
		(-1)^{\frac{k-k'-1}{2}}2^{q-\frac{k+k'-1}{2}}
		\frac{\sqrt{(2k+1)(2k'+1)}}{(k-k')(k+k'+1)}\\
		\qquad\times\sqrt{\frac{(k-q)!(k'-q)!}{(k+q)!(k'+q)!}}\,
		\frac{(k'+q)!!(k+q-1)!!}{(\frac{k'-q-1}{2})!(\frac{k-q}{2})!}\\
		\qquad\qquad \mathrm{if}\ k-q\ \mathrm{even\ and}\ k'-q\ \mathrm{odd}\\
		(-1)^{\frac{k-k'-1}{2}}2^{q-\frac{k+k'-1}{2}}
		\frac{\sqrt{(2k+1)(2k'+1)}}{(k-k')(k+k'+1)}\\
		\qquad\times\sqrt{\frac{(k-q)!(k'-q)!}{(k+q)!(k'+q)!}}\,
		\frac{(k+q)!!(k'+q-1)!!}{(\frac{k-q-1}{2})!(\frac{k'-q}{2})!}\\
		\qquad\qquad \mathrm{if}\ k'-q\ \mathrm{even\ and}\ k-q\ \mathrm{odd}\\
		0\ \mathrm{otherwise.}
	}
\end{eqnarray}

\section*{References}
\bibliography{MPQ}

\begin{thebibliography}{10}

\bibitem{QuantumStateEstimation}
M.~Paris and J.~{\v R}eh{\'a}{\v c}ek, editors.
\newblock {\em Quantum State Estimation}, volume 649 of {\em Lect. Notes Phys.}
\newblock Springer, Berlin Heidelberg, 2004.

\bibitem{Treutlein2004}
P.~Treutlein, P.~Hommelhoff, T.~Steinmetz, T.~W. H\"ansch, and J.~Reichel.
\newblock Coherence in microchip traps.
\newblock {\em Phys. Rev. Lett.}, 92(20):203005, 2004.

\bibitem{Esteve2008}
J.~Est{\`e}ve, C.~Gross, A.~Weller, S.~Giovanazzi, and M.~K. Oberthaler.
\newblock Squeezing and entanglement in a {B}ose--{E}instein condensate.
\newblock {\em Nature}, 455:1216--1219, 2008.

\bibitem{Appel2009}
J.~Appel, P.~J. Windpassinger, D.~Oblak, U.~B. Hoff, N.~Kj{\ae}rgaard, and
  E.~S. Polzik.
\newblock Mesoscopic atomic entanglement for precision measurements beyond the
  standard quantum limit.
\newblock {\em Proc. Nat. Acad. Sci.}, 106(27):10960--10965, 2010.

\bibitem{Gross2010}
C.~Gross, T.~Zibold, E.~Nicklas, J.~Est\`eve, and M.~K. Oberthaler.
\newblock Nonlinear atom interferometer surpasses classical precision limit.
\newblock {\em Nature}, 464:1165--1169, 2010.

\bibitem{Riedel2010}
M.~F. Riedel, P.~B\"ohi, Y.~Li, T.~W. H\"ansch, A.~Sinatra, and P.~Treutlein.
\newblock Atom-chip-based generation of entanglement for quantum metrology.
\newblock {\em Nature}, 464:1170--1173, 2010.

\bibitem{Leroux2010b}
I.~D. Leroux, M.~H. Schleier-Smith, and V.~Vuleti\'c.
\newblock Implementation of cavity squeezing of a collective atomic spin.
\newblock {\em Phys. Rev. Lett.}, 104:073602, 2010.

\bibitem{SchleierSmith2010}
M.~H. Schleier-Smith, I.~D. Leroux, and V.~Vuleti\'c.
\newblock States of an ensemble of two-level atoms with reduced quantum
  uncertainty.
\newblock {\em Phys. Rev. Lett.}, 104:073604, 2010.

\bibitem{Leroux2010a}
I.~D. Leroux, M.~H. Schleier-Smith, and V.~Vuleti\'c.
\newblock Orientation-dependent entanglement lifetime in a squeezed atomic
  clock.
\newblock {\em Phys. Rev. Lett.}, 104:250801, 2010.

\bibitem{Friedenauer2008}
A.~Friedenauer, H.~Schmitz, J.~T. Glueckert, D.~Porras, and T.~Schaetz.
\newblock Simulating a quantum magnet with trapped ions.
\newblock {\em Nature Physics}, 4:757--761, 2008.

\bibitem{Sackett2000}
C.~A. Sackett, D.~Kielpinski, B.~E. King, C.~Langer, V.~Meyer, C.~J. Myatt,
  M.~A. Rowe, Q.~A. Turchette, W.~M. Itano, D.~J. Wineland, and C.~Monroe.
\newblock Experimental entanglement of four particles.
\newblock {\em Nature}, 404:256--259, 2000.

\bibitem{Leibfried2005}
D.~Leibfried, E.~Knill, S.~Seidelin, J.~Britton, R.~B. Blakestad,
  J.~Chiaverini, D.~B. Hume, W.~M. Itano, J.~D. Jost, C.~Langer, R.~Ozeri,
  R.~Reichle, and D.~J. Wineland.
\newblock Creation of a six-atom `{S}chr{\"o}dinger cat' state.
\newblock {\em Nature}, 438:639--642, 2005.

\bibitem{Haeffner2005}
H.~H{\"a}ffner, W.~H{\"a}nsel, C.~F. Roos, J.~Benhelm, D.~Chek{-}al{-}kar,
  M.~Chwalla, T.~K{\"o}rber, U.~D. Rapol, M.~Riebe, P.~O. Schmidt, C.~Becher,
  O.~G{\"u}hne, W.~D{\"u}r, and R.~Blatt.
\newblock Scalable multiparticle entanglement of trapped ions.
\newblock {\em Nature}, 438:643--646, 2005.

\bibitem{Treutlein2006b}
P.~Treutlein, T.~Steinmetz, Y.~Colombe, B.~Lev, P.~Hommelhoff, J.~Reichel,
  M.~Greiner, O.~Mandel, A.~Widera, T.~Rom, I.~Bloch, and T.~W. H{\"a}nsch.
\newblock Quantum information processing in optical lattices and magnetic
  microtraps.
\newblock {\em Fortschr. Phys.}, 54(8-10):702--718, 2006.

\bibitem{Benhelm2008}
J.~Benhelm, G.~Kirchmair, C.~F. Roos, and R.~Blatt.
\newblock Towards fault-tolerant quantum computing with trapped ions.
\newblock {\em Nature Physics}, 4:463--466, 2008.

\bibitem{Newton1968}
R.~G. Newton and B.-l. Young.
\newblock Measurability of the spin density matrix.
\newblock {\em Annals of Physics}, 49(3):393--402, 1968.

\bibitem{Klose2001}
G.~Klose, G.~Smith, and P.~S. Jessen.
\newblock Measuring the quantum state of a large angular momentum.
\newblock {\em Phys. Rev. Lett.}, 86:4721--4724, 2001.

\bibitem{Buecker2009}
R.~B\"ucker, A.~Perrin, S.~Manz, T.~Betz, Ch. Koller, T.~Plisson, J.~Rottmann,
  T.~Schumm, and J.~Schmiedmayer.
\newblock Single-particle-sensitive imaging of freely propagating ultracold
  atoms.
\newblock {\em New J. Phys.}, 11:103039, 2009.

\bibitem{Ockeloen2010}
C.~F. Ockeloen, A.~F. Tauschinsky, R.~J.~C. Spreeuw, and S.~Whitlock.
\newblock Detection of small atom numbers through image processing.
\newblock {\em Phys. Rev. A}, 82:061606(R), 2010.

\bibitem{James2001}
D.~F.~V. James, P.~G. Kwiat, W.~J. Munro, and A.~G. White.
\newblock Measurement of qubits.
\newblock {\em Phys. Rev. A}, 64:052312, 2001.

\bibitem{Schleich}
W.~P. Schleich.
\newblock {\em Quantum Optics in Phase Space}.
\newblock Wiley-VCH, Berlin, 2001.

\bibitem{Dowling1994}
J.~P. Dowling, G.~S. Agarwal, and W.~P. Schleich.
\newblock Wigner distribution of a general angular-momentum state: Applications
  to a collection of two-level atoms.
\newblock {\em Phys. Rev. A}, 49(5):4101--4109, 1994.

\bibitem{Vogel1989}
K.~Vogel and H.~Risken.
\newblock Determination of quasiprobability distributions in terms of
  probability distributions for the rotated quadrature phase.
\newblock {\em Phys. Rev. A}, 40(5):2847--2849, 1989.

\bibitem{Smithey1993}
D.~T. Smithey, M.~Beck, M.~G. Raymer, and A.~Faridani.
\newblock Measurement of the {W}igner distribution and the density matrix of a
  light mode using optical homodyne tomography: Application to squeezed states
  and the vacuum.
\newblock {\em Phys. Rev. Lett.}, 70(9):1244--1247, 1993.

\bibitem{Breitenbach1997}
G.~Breitenbach, S.~Schiller, and J.~Mlynek.
\newblock Measurement of the quantum states of squeezed light.
\newblock {\em Nature}, 387:471--475, 1997.

\bibitem{Agarwal1998}
G.~S. Agarwal.
\newblock State reconstruction for a collection of two-level systems.
\newblock {\em Phys. Rev. A}, 57(1):671--673, 1998.

\bibitem{Gerlach1922}
W.~Gerlach and O.~Stern.
\newblock Der experimentelle {N}achweis der {R}ichtungsquantelung im
  {M}agnetfeld.
\newblock {\em Z. Phys.}, 9(1):349--352, 1922.

\bibitem{Zare}
R.~N. Zare.
\newblock {\em Angular Momentum}.
\newblock Wiley-Interscience, 1988.

\bibitem{KakSlaney}
A.~C. Kak and M.~Slaney.
\newblock {\em Principles of Computerized Tomographic Images}.
\newblock Classics in Applied Mathematics. IEEE Press, 1988.

\bibitem{Hinshaw2003b}
G.~Hinshaw, D.~N. Spergel, L.~Verde, R.~S. Hill, S.~S. Meyer, C.~Barnes, C.~L.
  Bennett, M.~Halpern, N.~Jarosik, A.~Kogut, E.~Komatsu, M.~Limon, L.~Page,
  G.~S. Tucker, J.~L. Weiland, E.~Wollack, and E.~L. Wright.
\newblock First-year {W}ilkinson microwave anisotropy probe ({WMAP})
  observations: The angular power spectrum.
\newblock {\em Astrophys. J. Supp. Series}, 148:135--159, 2003.

\bibitem{Banaszek2000}
K.~Banaszek, G.~M. D'{A}riano, M.~G.~A. Paris, and M.~F. Sacchi.
\newblock Maximum-likelihood estimation of the density matrix.
\newblock {\em Phys. Rev. A}, 61:010304(R), 2000.

\bibitem{Schulten1975}
K.~Schulten and R.~G. Gordon.
\newblock Exact recursive evaluation of $3j$- and $6j$-coefficients for
  quantum-mechanical coupling of angular momenta.
\newblock {\em J. Math. Phys.}, 16(10):1961--1970, 1975.

\bibitem{Ashour1964}
A.~A. Ashour.
\newblock On some formulae for integrals of associated {L}egendre functions.
\newblock {\em Quart. Journ. Mech. and Applied Math.}, 17(4):513--523, 1964.

\end{thebibliography}

\end{document}